\documentclass[aps,prl,doublecol,figures]{epl2}
\usepackage[utf8]{inputenc} 
\usepackage[T1]{fontenc}    %
\usepackage[english]{babel} 
\usepackage{amssymb,amsmath,mathtools} 
\usepackage{graphicx} 
\usepackage{xcolor}

\usepackage{hyperref}   

\begin{document}

\title{From Integrable to Chaotic Systems: Universal Local Statistics of Lyapunov exponents}

\author{Gernot Akemann \inst{1}  \and Zdzislaw Burda \inst{2} \and Mario Kieburg \inst{1}}

\institute{
\inst{1} Faculty of Physics, Bielefeld University, Postfach 100131, D-33501 Bielefeld, Germany\\
\inst{2} Faculty of Physics and Applied Computer Science,
AGH University of Science and Technology, al. Mickiewicza 30, PL-30059 Krakow, Poland
}

\abstract{
Systems where time evolution follows a multiplicative process are ubiquitous in physics. We study a toy model  
for such systems where each time step is given by multiplication with an independent random
$N\times N$ matrix with complex Gaussian elements, the complex Ginibre ensemble. 
This model allows to explicitly compute the Lyapunov exponents and local correlations amongst them, when the number of factors $M$ becomes large. While the smallest eigenvalues always remain deterministic, which is {also}
the case for many chaotic quantum systems, 
we identify a critical double scaling limit $N\sim M$ for the rest of the spectrum.
It interpolates between the known deterministic behaviour of the Lyapunov
exponents for $M\gg N$ (or $N$ fixed) and universal random matrix statistics
for $M\ll N$ (or $M$ fixed), characterising chaotic behaviour.
After unfolding this agrees with Dyson’s Brownian Motion starting from equidistant positions
in the bulk and at the soft edge of the spectrum.
This universality statement is further corroborated by numerical experiments,
multiplying different kinds of random matrices.
It leads us to conjecture a much wider applicability in
complex systems, that display a transition from deterministic to chaotic behaviour. 
}

\maketitle

\section{Introduction}

Random matrices have a long tradition in describing universal aspects of
spectral statistics, {typically of static Hamiltonians} in quantum physics~\cite{unfold}.
Today's  applications of random matrix theory (RMT)
cover all areas of physics and include other sciences, cf.
\cite{handbook} for a recent compilation.
The random matrix description is usually based
on global symmetries and not on dynamical principles.
However, already in early studies~\cite{Dyson} Dyson analysed
dynamical aspects of Brownian motion (BM) in matrix space,
{especially the dynamics which are formed by sums of matrices}. This work
initiated a discussion also on random matrix dynamics, stochastic processes in matrix space
and the underlying dynamical principles.

A different and \textit{a priori} unrelated idea to include dynamics is to study
systems where the time dependence is multiplicative.
{This is the main objective of the present work.}
 Such processes can be
found in many physical systems e.g. in the form of transfer matrices, leading
to the 
Dorokhov–Mello–Pereyra–Kumar (DMPK) 
equation~\cite{D,MPK}, through propagation kernels~\cite{jw,blBook}, or more recently in quantum entanglement~\cite{cnz}.
Long time ago Furstenberg and Kesten~\cite{fk} have proposed to multiply
$M$ 
random matrices  as a toy model for chaotic dynamical systems. The corresponding Lyapunov exponents
become deterministic in the large-$M$ limit~\cite{n,ni}, and we refer to the
reviews~\cite{cpvBook} for applications in disordered systems and to~\cite{Viana}
for the mathematical statements. It was conjectured early in~\cite{KlausFrahm}
that the  Lyapunov spectrum and the universal 
statistics of a
single random matrix are related. The claim was substantiated in a stability
analysis of complex systems using stochastic processes~\cite{IpsenSchomerus}. 
Different recent applications of such multiplicative processes
include non-perturbative quantum
gravity with space-time foliation~\cite{ajl}, 
or the propagation of information in telecommunication~\cite{m,VerduTulino}. Multiplying $M$ random $N\times  N$ matrices
applies further to  multi-layered complex networks with $M$ layers and 
$N$ degrees of freedom per layer~\cite{BBCGGRSWZ}, 
to the  complexity of random maps~\cite{if},
and in  computer science (cf.~\cite{cknBook}) through machine learning~\cite{l}. All these examples can be viewed as certain realisations of progressive scattering, where a distinguished (time) direction exists.
Therefore, the product matrix can be often interpreted
as a transfer matrix. We will stick to this interpretation in
our discussion.

{How can one relate time dependent processes and universal equilibrium statistics of RMT? The standard dichotomy considered in RMT for the spacing distribution of consecutive energy levels of a static Hamiltonian is between Poisson statistics characterising integrable systems - the Berry--Tabor conjecture~\cite{BT} - and the (approximate) Wigner surmise in RMT characterising chaotic quantum systems - the Bohigas--Giannoni--Schmit conjecture~\cite{BGS}.
Such transitions have been an intense object of study~\cite{Stoeckmann}.}

{Fixing all quantum numbers but one in an integrable Hamiltonian quantum system, like the hydrogen atom or harmonic oscillator, the respective spectrum can be unfolded to become equidistant. It is the incommensurable superposition of several such picket fence spectra that then becomes Poisson~\cite{BGS}. In contrast, our results only exhibit a single picket fence structure, realised by the individual Lyapunov exponents, in the limit where they become deterministic.
When increasing the system size compared to the number of time steps, the Lyapunov exponents start to interact with each other. We will find that their statistics can then be described by Dyson’s BM with fixed initial conditions.}

{Rather than focussing on the level spacing distribution which is a classical measure for random matrix statistics, we focus on the correlation kernel, encoding all correlation functions in the underlying determinantal point process~\cite{Kurt}.
For recent work on the spacing of Lyapunov exponents, cf.~\cite{Hanada}.}

Let us review some recent developments in products of random matrices,
cf.~\cite{ipsenakemann}. Multiplying $M$ independent complex Ginibre matrices~\cite{g} having $N^2$ complex normal elements, 
exact analytical expressions for the joint probability distribution functions 
of the eigenvalues~\cite{ab} and singular values~\cite{akw,aik}
have been derived for finite $M$ and $N$. They are given by 
a determinantal point process which means that all $k$-point correlation functions can be written as determinants,
\begin{equation}\label{kpoint}
R_k(x_1,\ldots,x_k)=\det[K(x_j,x_l)]_{j,l=1,\ldots,k}\ ,
\end{equation}
of a correlation kernel $K(x_j,x_l)$. The kernel is expressed in terms of
Meijer G-functions~\cite{NIST}. {The level density $R_1(x)=K(x,x)$ is the}
simplest example.
These findings have initiated considerable activity
for finite $M$ and $N\rightarrow \infty$, see
e.g.~\cite{kuijlaars2,kuijlaars,lwz} and references in the
review~\cite{ipsenakemann} for more general products.

Based on these findings 
the full statistics for finite Lyapunov exponents was derived in~\cite{abk} at
fixed $M$ and $N$. Therein, it was argued that for the local statistics
a non-trivial double scaling limit $M,N \rightarrow \infty$ should exist,
leading away from the deterministic limit $M\to\infty$, for $N$ fixed (or
$N\to\infty$) \cite{n,ni}.
In contrast, for fixed $M$  the universal local statistics of a single
Gaussian Unitary Ensemble (GUE)~\cite{GUE} was found in the bulk and at the soft edge of
the spectrum of Lyapunov exponents~\cite{lwz}. 
{A similar} transition
from deterministic to GUE statistics
was also seen in~\cite{IpsenSchomerus} for the solution of the DMPK equation,
{ where the authors pointed out a connection to Dyson's BM.
Let us emphasise that the matrix products taken for DMPK are perturbations of the unit matrix
 while in the present work we consider more general products of complex matrices, for which this transition has not
 been studied before, and point out the general mechanism behind this transition.}

\section{Qualitative Discussion}

We consider $M+1$ time  steps (layers)
of the system, where each step  is described by an $N$-dimensional vector $\vec{v}_j$, $j=0,\ldots,M$.
In the simplest situation 
$\vec{v}_j=X_j\vec{v}_{j-1}$, the evolution from an initial condition $\vec{v}_0$ to time 
$M$  follows the product matrix {$Y=X_M\cdots X_1$. We call $Y$ the transfer matrix.} 
Much information is encoded in the Lyapunov exponents, being eigenvalues of the  
{\it Lyapunov matrix}
\begin{equation}
L = \frac{1}{2M} 
\log Y^\dagger Y= \frac{1}{2M}  
\log (X_M \cdots X_1)^\dagger (X_M \cdots X_1) ,
\label{eq:L}
\end{equation}
where the invertibility 
of the product is assumed throughout this work.

Let us consider an arbitrary product of operators,
without prior assumption  
about their dependence. 
While our explicit calculations are for Ginibre matrices, 
our discussion should apply to more general situations, as well. As an example, in Fig.~\ref{fig:sim.bulk}, we compare the microscopic level density in the bulk of the spectrum of the product of Ginibre matrices, complex Bernoulli matrices (with entries  $0,\pm1,\pm i,\pm 1\pm i$), and sums of Ginibre matrices $A_j$,  
$X_j=A_j+A_{j-1}$, with $A_0=0$. The latter choice can be understood as a model for progressive scattering with a short memory, making it non-Markovian. For $M=2$ it was considered in~\cite{akemannstrahov}.
{Indeed, we could also choose $X_j=1+\gamma\delta t+A_j\sqrt{\delta t}$
with $A_j$ independent Ginibre matrices, the scalar $\gamma$ being proportional to the variance of $A_j$, and the time increment $\delta t\propto 1/M$. This model leads to the DMPK equation~\cite{IpsenSchomerus} and in the limit $M\to\infty$  it models the Brownian stochastic process
\begin{equation}\label{stochproc}
dY(t)=[dA(t)+\gamma dt]Y(t).
\end{equation}
In the present work, we will not go into details of this process and refer to~\cite{IpsenSchomerus}.}

{First, we want to recall what is known.} In several works, e.g.~\cite{tutubalin,n,richards,reddy,kieburg}, it has been shown for various kinds of $X_j$ being i.i.d.  that the distributions of the Lyapunov exponents $\lambda_j$ become asymptotically Gaussian when choosing $M\gg N$. The level density is then given by
\begin{equation}
R_1(\lambda)\approx {\sum}_{j=1}^N
\exp\bigl[ -(\lambda-\bar{\lambda}_j)^2/(2 \sigma_j^2)\bigl]/\sqrt{2\pi \sigma_j^2},
\label{eq:gaussian}
\end{equation}
with deterministic means $\bar{\lambda}_j$~\cite{n} and standard deviations~$\sigma_j$. 
For {$M$ Ginibre matrices} they are given by~\cite{abk}
\begin{equation}
\bar{\lambda}_j =\psi(j)/2\ {\rm and}\ \sigma_j =\sqrt{ \psi'(j)/(4M)},
\label{eq:ls}
\end{equation}
where $\psi(z) = (\log \Gamma(z))'$ is the digamma function~\cite{NIST}.
When $N/M\to0$ the level density becomes a sum of Dirac delta functions, 
corresponding to picket fence statistics after proper unfolding.

Clearly, the approximation~\eqref{eq:gaussian} holds  as long as the overlap between individual Lyapunov exponents is small, meaning that the $j$th {\it width-to-spacing-ratio} (WSR) defined by
\begin{equation}
{\rm WSR}_j = \frac{1}{2}\frac{\sigma_j+\sigma_{j-1}}{\overset{\ }{\bar{\lambda}_j-\bar{\lambda}_{j-1}}},\ {\rm for\ Ginibre}\ {\rm WSR}_j \overset{j\gg1}{\approx}\sqrt{\frac{j}{M}},
\label{eq:wsr}
\end{equation}
is small. In the case of $M$ Ginibre matrices this is always satisfied 
when $j$ is fixed while $N,M\to\infty$ (hard edge scaling, 
smallest eigenvalue), 
regardless of any relation between $N$ and $M$.

\begin{figure}[t!]
\includegraphics[width=0.45\textwidth]{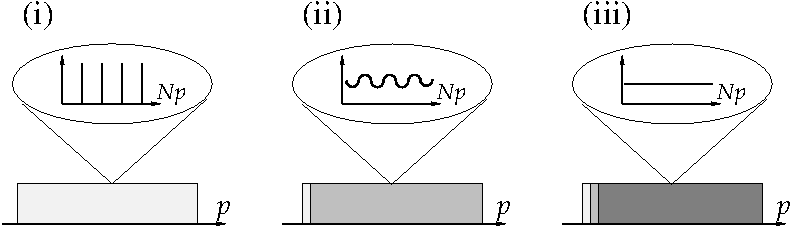}
\caption{Sketches of the unfolded microscopic spectra (insets) of the Lyapunov exponents in different double scaling limit:
(i) $M\gg N$, (ii) $M\propto N$, and (iii) $M\ll N$.
The ratio $p = j/N \in [0,1]$, labels the number $j$ of ordered eigenvalues.
Picket fence statistics is in light grey, the interpolating  
curve in grey, and GUE statistics in dark grey.
\label{fig:local_bulk}}
\end{figure} 

In contrast, for the largest, $N$th Lyapunov exponent the
${\rm WSR}_N \approx \sqrt{N/M} $ is only small when $M\gg N$. 
For $M\ll N$ we cannot expect the approximation~\eqref{eq:gaussian} to hold; the overlap between individual eigenvalue distributions becomes large, their interactions become important, and the standard GUE results~\cite{GUE} should follow. The critical regime is when $M\propto N$ or more precisely $M\propto j$.
In the ensuing sections, we will concentrate on this regime.

{Summarizing the discussion above,} 
we obtain the following picture illustrated in Fig.~\ref{fig:local_bulk} for the product of $M$ Ginibre matrices.
When studying the double scaling limit
$M, N \rightarrow \infty$ with the relation $a= \lim_{N\rightarrow \infty} N/M$, we find three distinct asymptotic regimes: (i) $M=M(N)$ increasing super-linearly ($a=0$), (ii) linearly ($0<a<\infty$),  or (iii) sub-linearly ($a=\infty$) with $N$.
 {This identification of scales is our first main result. It shows us that there is always a critical spectral regime for $M\ll N$ in contrast to standard random matrix models and more natural for physical operators.}
Below we will only consider the bulk and soft edge 
(largest eigenvalue)
of the Lyapunov exponents as they display new features.
The local WSR is parametrised by $a$ as
${\rm WSR}_{j=Np} = \sqrt{ap}$, see~\eqref{eq:wsr}. 
It increases when moving from left to right
in the spectrum, cf. Fig.~\ref{fig:local_bulk}. A similar picture is expected for a general product of operators, 
with the WSR as a parameter, cf. Fig.~\ref{fig:sim.bulk}.

\section{Unfolding for Ginibre}

To compare microscopic properties of different spectra one needs to unfold them via the averaged cumulative density~\cite{unfold}
\begin{equation}\label{quantile}
\bar{N}(\lambda)=\int_{-\infty}^\lambda\bar{R}_1(\lambda')d\lambda'.
\end{equation}
$\bar{R}_1(\lambda)$ is the non-fluctuating part of the level density,
approaching the macroscopic level density at $N\to\infty$.
When no analytical formulas for $\bar{R}_1(\lambda)$ and $\bar{N}(\lambda)$ are available, e.g. for comparison with experimental data, one has to perform a fit.
 {Fortunately, $\bar{R}_1(\lambda)$ is known 
analytically 
for products of Ginibre matrices~\cite{akw}. 
In the bulk of the spectrum, taking $M\to\infty$ regardless of whether and how $N$ approaches infinity, this density is $\bar{R}_1(\lambda)=2e^{2\lambda}\Theta(N-e^{2\lambda})$,  with $\Theta$ the Heaviside step function. 
This mapping will be explained in detail below. 
At the soft edge where $\bar{R}_1(\lambda)$ vanishes as a square root, we will specify the fit for the function $\bar{N}(\lambda)$.}

The eigenvalue statistics of the random matrix $\exp[2M L]$ 
follows a 
determinantal point process~\eqref{kpoint} at fixed $M$ and $N$ with the kernel
\begin{eqnarray}
K({x},{y}) &=& 1/x\, {\sum}_{j=1}^{N} \left(x/y\right)^j G_j({y}),
\label{eq:main1}\\
G_j({y}) &=&\int_{-i\infty}^{+i\infty} \frac{dt}{2\pi i}\frac{\sin (\pi t)}{\pi t}\ {y}^t 
\left(\frac{\Gamma(j-t)}{\Gamma(j)}\right)^{M+1}\nonumber\\
&&\times \frac{\Gamma(N-j+1+t)}{\Gamma(N-j+1)}.
\label{eq:main2}
\end{eqnarray}
Here, $G_j$ is essentially a Meijer G-function,
 {and Eq.~\eqref{eq:main1} is} equivalent to the kernel derived in~\cite{aik}.
The normalized level density of $L$ is
\begin{equation}
\begin{split}
\frac{R_1(\lambda)}{N} =& \frac{2M}{N}e^{2M \lambda} K(e^{2M \lambda},e^{2M \lambda})= \frac{2M}{N} \sum_{j=1}^{N} G_j(e^{2M \lambda}).
\end{split}
\label{eq:c1}
\end{equation}
The prefactor stems from the Jacobian of the unfolding. Similarly, the $k$-point correlation functions~\eqref{kpoint} of $L$ are given by the kernel $2Me^{2M \lambda_j} K(e^{2M \lambda_j},e^{2M \lambda_l}) $.

Due to the explicit form of the kernel~\eqref{eq:main1} we can take various limits. 
For fixed $N$ and sufficiently large $M$ one reproduces the Gaussian result~\eqref{eq:gaussian}: For $y=e^{2 M\lambda}$, the integral~\eqref{eq:main2}  picks up its main contribution from the saddle point at $t=0$.
When $M\rightarrow \infty$ with finite $j$, this yields a Dirac delta 
at $\bar{\lambda}_j=\psi(j)/2$. 
Thence, the spectrum at the hard edge is always discrete and deterministic on a microscopic scale and, after proper unfolding, 
gives picket fence statistics.

The unfolding in the bulk follows from the asymptotic behaviour of the eigenvalues of the matrix $e^{2L}/N$ which are approximately $e^{\psi(j)}/N$. For
\begin{equation}
j=Np\ {\rm with}\ p\in(0,1)
\end{equation}
 being of order $N$,  
the limiting form is $e^{\psi(j)}/N \approx p$ for $N\rightarrow \infty$, following from the asymptotic formula $\psi(Np) \approx \log(N)+\log(p) + \mathcal{O}(1/N)$.
Thus, the eigenvalues $p = e^{2\lambda}/N$ 
of the matrix $\exp[2L]/N$ 
are uniformly distributed on $(0,1)$. This only holds when being away from the edges at $p=0,1$. The map $\lambda\mapsto p$ is just the cumulative distribution~\eqref{quantile} (quantile) and therefore the proper unfolding. 

\section{Bulk Statistics}

{Zooming into the microscopic scale in the bulk at $p\in(0,1)$, we consider  two neighbouring points
\begin{equation}
x=(Np +\xi)^M\ {\rm and}\ y=(Np +\zeta)^M
\end{equation}
 in the kernel~\eqref{eq:main1}, with $\xi$ and $\zeta$ of order unity.
 Multiplied by the Jacobian $M(Np+\xi)^{M-1}$, the  kernel~\eqref{eq:main1} in the bulk becomes
\begin{equation}
\label{eq:kernel_bulk}
\begin{split}
K_{\rm bulk}\left(\xi,\zeta;a\right)=&\lim_{N,M\to\infty}M(Np+\xi)^{M-1}\frac{g(\xi)}{g(\zeta)}\\
&\times K((Np+ \xi + \Delta p)^M, (Np+ \zeta + \Delta p)^M)\\
=&\frac{1}{2\pi a p}{\sum}_{j = -\infty}^{+\infty} e^{j(\xi-\zeta)/(ap)}\\
&\times\mathrm{Re}\biggl(\mathrm{erfi}\left[\frac{\pi}{2}\sqrt{2ap} + \frac{i}{\sqrt{2ap}} \left(
\zeta  - j \right)\right]\biggl),
\end{split}
\end{equation}
for $a=\lim_{N,M\to\infty}N/M\in(0,\infty)$.  
 {Here, $g(\xi)$ is an appropriate function to guarantee the existence of the limit, in this case 
$g(\xi)=(pN+\xi)^{Nj_0/a}$, with $j_0=[Np]$.
Because it drops out from the determiant that yields the $k$-point correlation functions~\eqref{kpoint}, we will not specify it any more below.}

 {In~\cite{ABK2019}, we give details of the derivation of~\eqref{eq:kernel_bulk}. To sketch the idea, we apply the saddle point approximation and expand the summand and  integrand in~\eqref{eq:main1} and~\eqref{eq:main2} about the summation index $j=Np+\delta j$ and the integration variable $t=i\delta t$, with $\delta j$ and $\delta t$ being of order one. This expansion is cumbersome but it is inevitable to go up to subleading orders since the leading orders cancel.}

Equation~\eqref{eq:kernel_bulk} is our second main result.
The imaginary error function is defined as $\mathrm{erfi}(z) = \frac{2}{\sqrt{\pi}}\int_0^z \exp[t^2] dt$. The local shift $\Delta p=Np-[Np]+1/2-ap\log[(1-p)/p]$ is model dependent and thus not universal. It tells us how far away from the hard and soft edge we zoom into the spectrum and what the distance of $Np$ to the closest integer $[Np]$ is.

We would like to point out that the limit~\eqref{eq:kernel_bulk} can be also found in the spectrum when $M$ behaves sub-linearly with $N$ ($a=\infty$). There is always a small transitional regime close to the hard edge limit where the index $j$ of the Lyapunov exponent is of the order $M$, see Fig.~\ref{fig:local_bulk}.

\begin{figure}[t!]
	\centering 
	\includegraphics[width=0.47\textwidth]{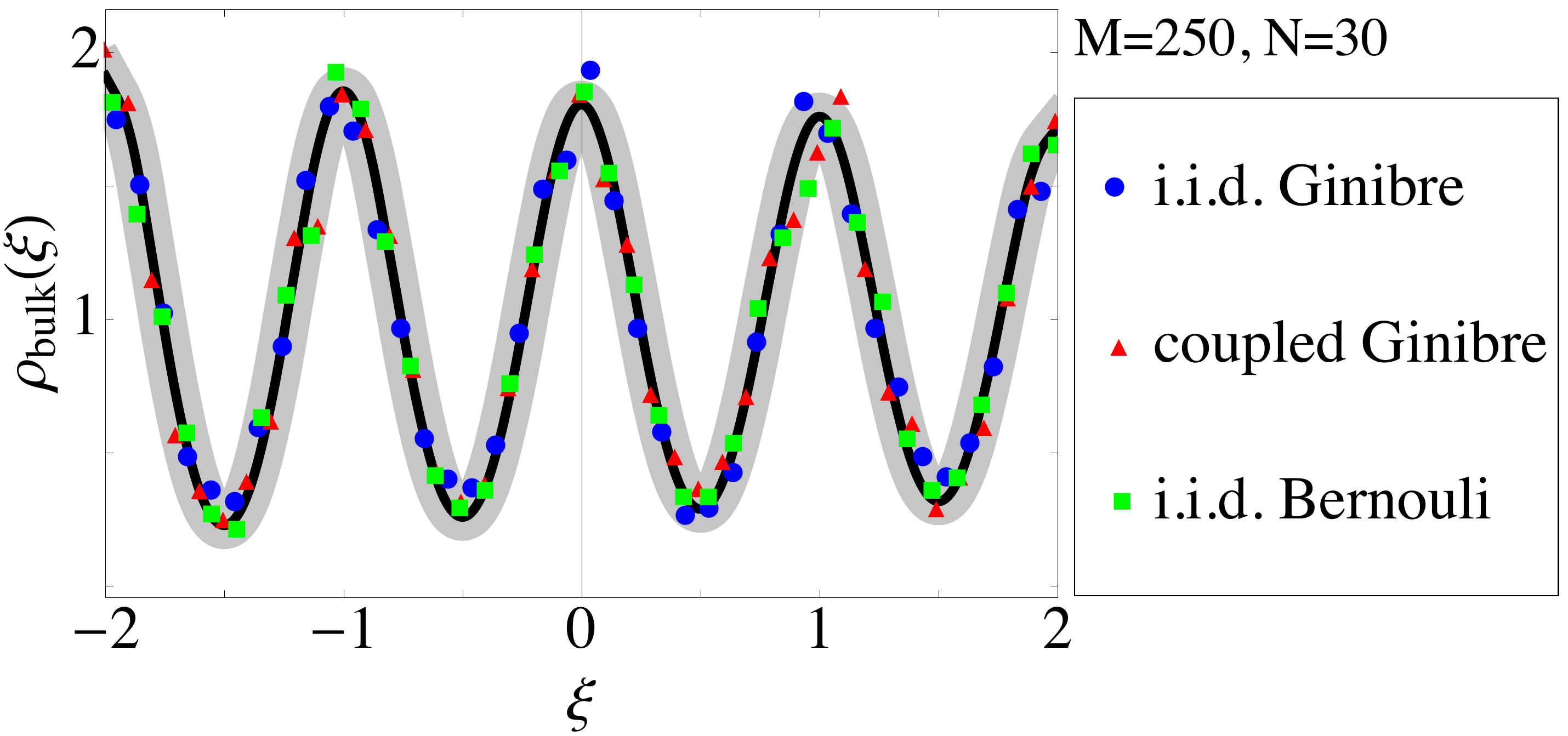}
	\caption{\label{fig:sim.bulk} Comparison of Monte Carlo simulations (symbols) of three matrix products introduced after~\eqref{eq:L}, at $M=250,\ N=30$, with the analytical result~\eqref{eq:ldb} 
	at ${\rm WSR}_{j=15}\approx0.24$ (black curve). Each product ensembles comprises $1192$ configurations, with a residual statistical error of about $10\%$ (grey area). Because of the small matrix dimension $N=30$,
	we expect further deviations for  the chosen eigenvalues $j=13,\ldots,17$ .
	To improve the fit we replaced $p=15/30=0.5$ by $p=(15+\xi)/30$ in~\eqref{eq:ldb}.
	}
\end{figure}

Let us reformulate~\eqref{eq:kernel_bulk} by the Poisson summation formula to
\begin{equation}
\label{eq:kernel_bulk.b}
\begin{split}
&K_{\rm bulk}\left(\xi,\zeta;a\right)=\frac{1}{\pi}e^{(\xi^2-\zeta^2)/(2ap)}\\
\times&{\rm Re}\biggl({\sum}_{n=-\infty}^\infty\frac{e^{-2\pi^2apn(n-1)+i\pi (\zeta+(2n-1)\xi)}}{[2\pi apn+i(\zeta-\xi)]}\biggl) .
\end{split}
\end{equation}
The prefactor $\exp[(\xi^2-\zeta^2)/(2ap)]$ can be skipped in the last formula
since it drops out in all correlation functions~\eqref{kpoint}, too. Though the second line of~\eqref{eq:kernel_bulk} has the advantage that each summand can be associated to a single Lyapunov exponent, the representation~\eqref{eq:kernel_bulk.b} reveals the intimate relation with the kernel from~\cite[Theorem 2.5]{KurtBrown}, 
where Dyson's BM with equidistant initial conditions 
has been studied, which is also a  
determinantal point process.
This agreement is quite surprising, and analytically
 it shows the universality of our limit~\eqref{eq:kernel_bulk}.
In~\cite{IpsenSchomerus} this particular BM~\cite{KurtBrown} has
 been identified with the solution of the DMPK equation.
Thus, it can be expected that our result~\eqref{eq:kernel_bulk} 
corresponds to the local spectral statistics of the Anderson
transition~\cite{D,MPK,KlausFrahm,Beenakker,IpsenSchomerus}.

\begin{figure}[t!]
	\centering 
	\includegraphics[width=0.47\textwidth]{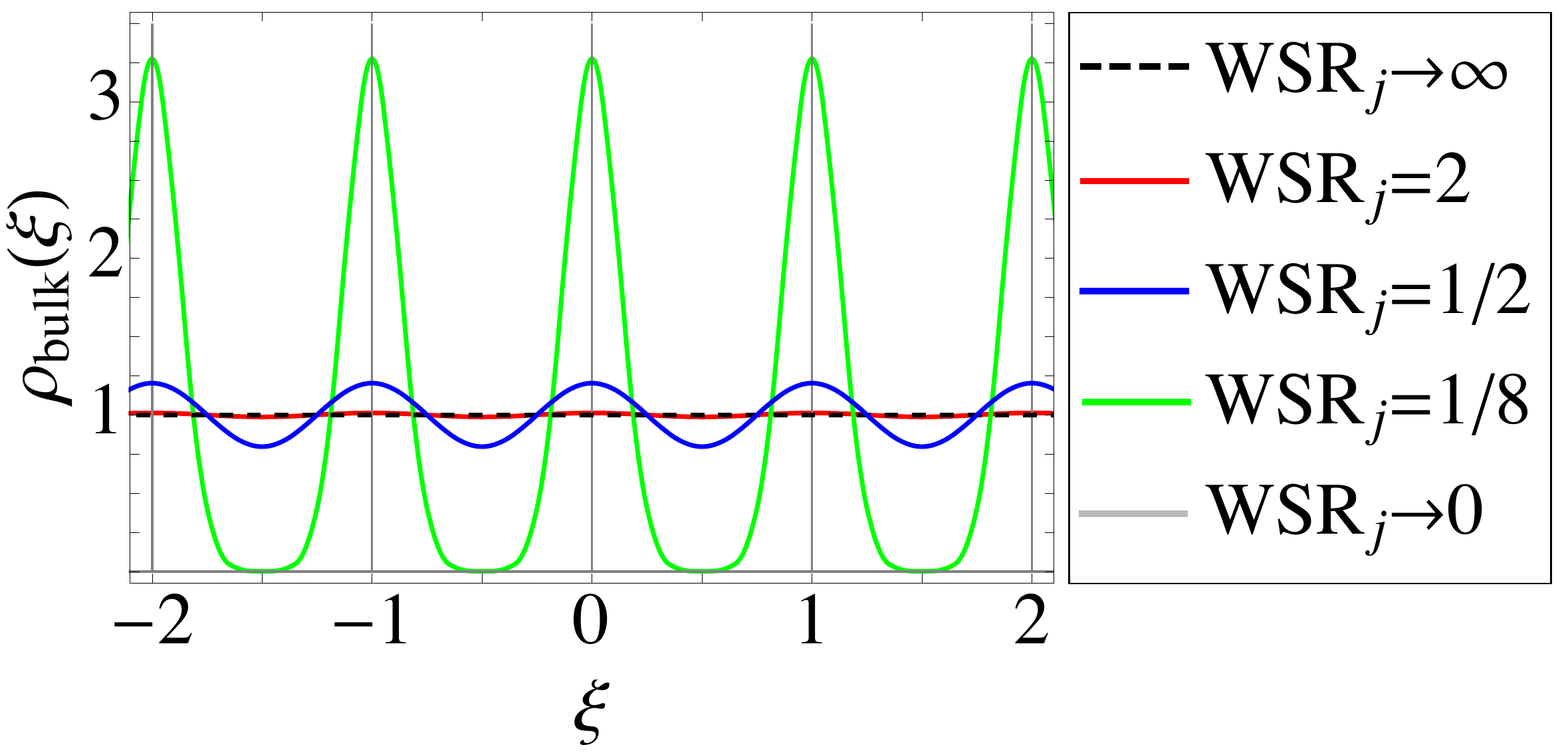}
	\caption{\label{fig:micro.dens.bulk} The unfolded microscopic level density~\eqref{eq:ldb} in the bulk for various values of ${\rm WSR}_j=\sqrt{j/N}=\sqrt{ap}$.
	The origin represents the point $Np$ where we zoom in.
The limiting constant density (dashed line) corresponds to the sine-kernel.
	}
\end{figure}

Let us discuss now the microscopic level density 
on the scale of the local mean level spacing, 
given by the relation
\begin{equation}
\rho_{\rm bulk}(\xi;a) = K_{\rm bulk}\left(\xi,\xi;a\right).
\label{eq:ldb}
\end{equation}
The behaviour of $\rho_{\rm bulk}(\xi;a) $ is shown in Fig.~\ref{fig:micro.dens.bulk} for various values of $a$. In order to further corroborate the universality found above, a
 comparison to Monte Carlo simulations of  three different products of random matrices 
is made in Fig.~\ref{fig:sim.bulk}. 

The interpolating microscopic density in the bulk~\eqref{eq:ldb} as well as the corresponding kernel~\eqref{eq:kernel_bulk} enjoy a discrete 
periodicity 
$\rho_{\rm bulk}(\xi;a)=\rho_{\rm bulk}(\xi+1;a)$
on a local scale.
A full continuous translation invariance \cite{lwz} is restored only in the limit $a\rightarrow \infty$ where
we analytically recover the sine-kernel~\cite{GUE} 
\begin{equation}
K_{\sin}\left(\xi,\zeta\right)= \frac{\sin[\pi (\xi-\zeta)]}{\pi (\xi-\zeta)}=\lim_{a\to\infty} K_{\rm bulk}\left(\xi,\zeta;a\right) .
\label{eq:sk}
\end{equation}
This 
follows from approximating the sum~\eqref{eq:kernel_bulk} by an integral that can be performed~\cite{ABK2019}. 
 Yet, there is {always}
a small region between the hard edge and the bulk, even for $a\to0$, where the transition kernel~\eqref{eq:ldb} can be still found because ${\rm WSR}_j$, see~\eqref{eq:wsr}, or equivalently $ap=j/M={\rm WSR}_j^2$ enters 
the kernel~\eqref{eq:ldb}.

For $a \rightarrow 0$ the
density~\eqref{eq:ldb} reduces to a sum of Dirac delta functions,
\begin{equation}
\lim_{a\to0}\rho_{\rm bulk}(\xi;a) = \sum_{j=-\infty}^\infty \delta\left(\xi- j\right),
\end{equation}
since the erfi-function narrows to a Gaussian. The same applies to any $k$-point correlation function, yielding picket fence statistics~\cite{BGS}.

\section{Soft Edge Statistics}

At the soft edge ($p=1$) the mean level spacing has a different scaling behaviour. 
Here, we adopt the finite $M$ scaling of~\cite{lwz} and find the position of the upper spectral edge of $\exp[2M L]$ at $N^M (M+1)^{M+1}/M^M$, when $N\rightarrow \infty$. The probability of finding an eigenvalue above the edge drops exponentially. 
Therefore, we zoom into the spectrum as
\begin{equation}
{x}=(M+1)^{M+1} a^M\left( 1 + a^{-2/3}\xi/M\right)^M .
\label{eq:Ss}
\end{equation}
The prefactor $a^{-2/3}$ in front of $\xi$ is reminiscent to the Airy-kernel scaling~\cite{Peter}. We insert this scaling into~\eqref{eq:main1} and fix the limit $a=\lim_{N,M\to\infty}N/M$. 
{After relabelling the summation index $j\to N-j$ and expanding the summand as well as integrand in $1/N\sim1/M$ in~\eqref{eq:main1} and~\eqref{eq:main2},}
the double scaling limit
 $M,N\to\infty$ \eqref{eq:Ss} leads to the interpolating kernel,
\begin{equation}
\label{eq:kernel_soft}
\begin{split}
&K_{\rm soft}\left(\xi,\zeta;a\right)\\
=&\lim_{N,M\to\infty}\frac{1}{a^{2/3}}\left( 1 + \frac{\xi}{M a^{2/3}}\right)^{M-1}\frac{\tilde{g}(\xi)}{\tilde{g}(\zeta)}K\left({x},{y}\right)\\
=&\int_{-\infty}^\infty \frac{dt}{2\pi  a^{2/3}}
\frac{\left(1 - \exp[-i t/a-(\xi-\zeta)/a^{2/3}]\right)^{i t-1}}{\Gamma(1+i t)} \\
&\times\exp\left[-\frac{t^2}{a} - i t \left(1 - \log a +\frac{1}{2a} +\zeta/a^{2/3}\right)\right] ,
\end{split}
\end{equation}
our third main result.  {Again we postpone the detailed derivation to~\cite{ABK2019}. The sum could be evaluated explicitly after the expansion.}

An equivalent result was derived independently in~\cite{LWW}.
The interpolating microscopic level density at the soft-edge reads $\rho_{\rm soft}(\xi;a)=K_{\rm soft}\left(\xi,\xi;a\right)$. 
We also find that the soft edge kernel~\eqref{eq:kernel_soft} appears in Dyson's BM~\cite{KurtBrown}, and via the identification~\cite{IpsenSchomerus}  also for the
DMPK equation.
Therefore, also the interpolating kernel at the soft edge is universal 
 and should 
 hold for more general products of operators, as long as a soft edge is present.

For $a\to\infty$, we recover the GUE Airy-kernel~\cite{Peter},
\begin{equation}
\begin{split}
&K_{\rm Airy}(\xi,\zeta)=\lim_{a\rightarrow \infty} \frac{\hat{g}(\xi)}{\hat{g}(\zeta)}K_{\rm soft}(\xi,\zeta;a)\\ 
&=
2^{-1/3}
\frac{{\rm Ai}(2^{1/3} \xi){\rm Ai}'(2^{1/3} \zeta)-{\rm Ai}(2^{1/3} \zeta){\rm Ai}'(2^{1/3} \xi)}{\xi-\zeta},
\end{split}
\label{eq:kernel_inf}
\end{equation}
depending on the Airy function. 
For the interpolating  microscopic density at the soft edge 
this implies
\begin{equation}
\lim_{a\rightarrow \infty} \rho_{\rm soft}(\xi;a) =
2^{1/3}{{\rm Ai}'}^2(2^{1/3} \xi)-2^{2/3} \xi {\rm Ai}^2(2^{1/3} \xi)\ .
\label{eq:se_inf}
\end{equation}

\begin{figure}[t!]
	\centering 
	\includegraphics[width=0.47\textwidth]{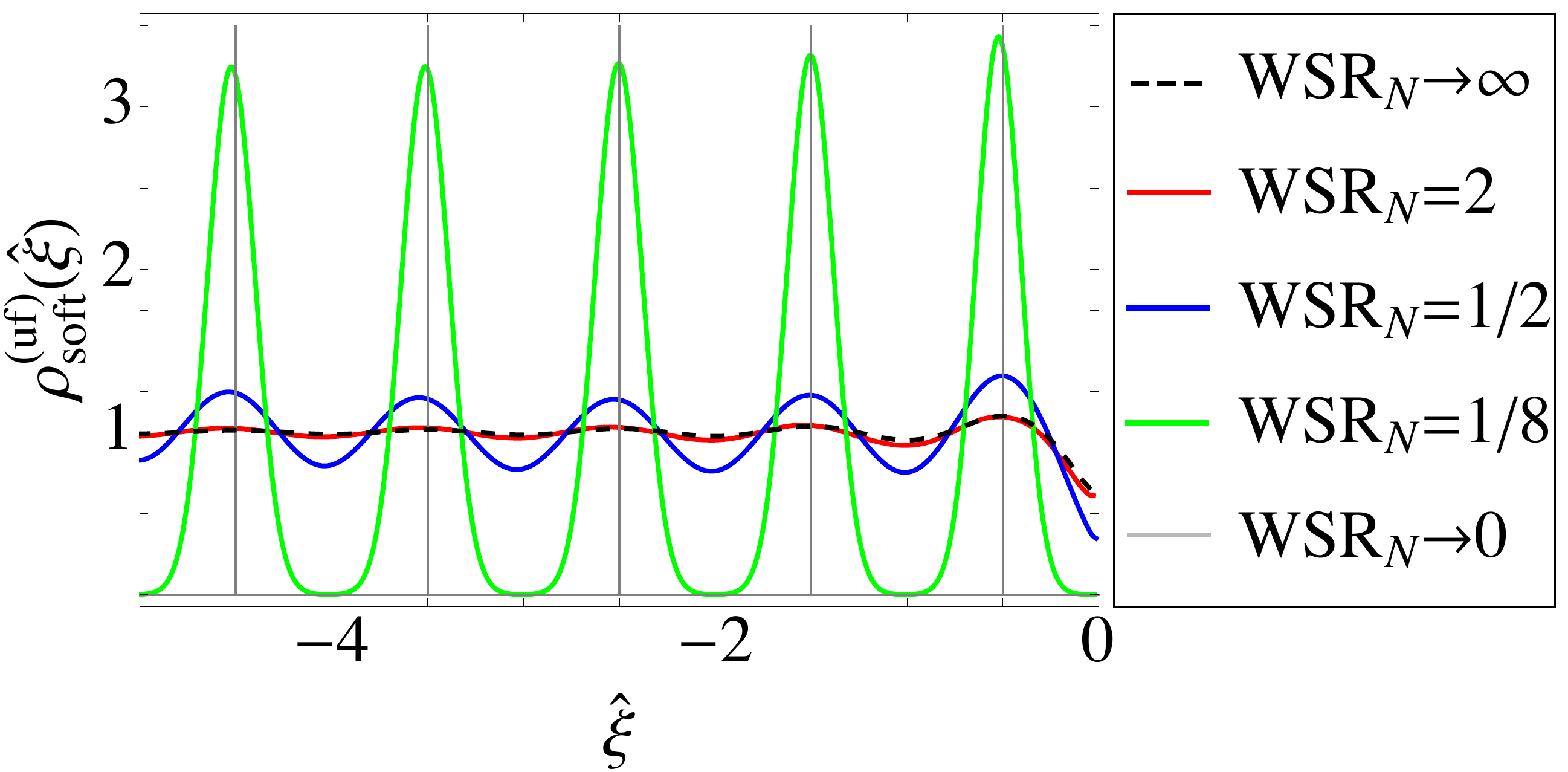}
	\caption{\label{fig:micro.dens.soft}
	The interpolating microscopic level density at the soft edge from~\eqref{eq:kernel_soft} after unfolding (uf),  for various values of ${\rm WSR}_{N}=\sqrt{a}$, with the upper bound of the macroscopic density at zero.
	After unfolding the limiting Airy-density~\eqref{eq:se_inf} is only weakly oscillating (dashed line).
	}
\end{figure}

As for the Airy-kernel~\eqref{eq:kernel_inf}, the spectrum of~\eqref{eq:kernel_soft} is 
not 
unfolded. 
For example, for large 
negative
argument $\xi$, the limiting Airy-density~\eqref{eq:se_inf} increases as $\sqrt{|\xi|}$ to the left, originating from the macroscopic semi-circular law of the GUE.
In Fig.~\ref{fig:micro.dens.soft} we compare the unfolded density for the interpolating kernel~\eqref{eq:kernel_soft} and the Airy-density. 
 {Our unfolding, fitting\eqref{quantile} by $\bar{N}(x)=ax+bx^{3/2}+cx^2$,
 is only valid inside the support of the macroscopic level density
which vanishes as a square root, cf. \cite{aik}.}
For the tails, which include the Tracy--Widom distribution~\cite{TW} for 
the GUE, a different means of comparison has to be sought.

In the opposite limit, for $a\to0$, 
we rescale $\xi=\hat{\xi}/a^{1/3}$,
\begin{equation}
\lim_{a\rightarrow 0} a^{-1/3}\rho_{\rm soft}(a^{-1/3}\hat{\xi};a) =
{\sum}_{j=1}^\infty \delta(\hat{\xi} + j - 1/2).
\label{eq:se0}
\end{equation}
This is the picket fence density {with a lower bound
and agrees with} the spectrum of a quantum harmonic oscillator. The rescaling by $a^{-1/3}$ is essential to get a normalized mean level spacing.

{Comparing the observations above with the largest eigenvalue distribution, it is well known that the one corresponding to} the Airy-density~\eqref{eq:se_inf} is given by the Tracy--Widom distribution~\cite{TW}. In contrast, in the picket fence limit 
it is approximately Gaussian,  see~\eqref{eq:gaussian}.
This was also found in~\cite{IpsenSchomerus} via a Fokker-Planck 
equation approach, and in~\cite{LWW}. The interpolating kernel~\eqref{eq:kernel_soft} thus describes an interpolation between the two, demanding further investigations.

\section{Conclusions}

The local statistical properties of Lyapunov exponents of {transfer matrices, modelled by products of random matrices,} have been discussed  in the limit $N,M\to\infty$, where $M$ typically represents the number of time steps or layers, and $N$ the degrees of freedom and thus the complexity of the underlying system. The 
critical scaling $a\propto N/M$ separates two different phases
{that can be}  associated with integrable or chaotic behaviour. 
For  $M\gg N$ the entire 
Lyapunov spectrum is deterministic. 
In the opposite limit $M\ll N$, the bulk and the largest Lyapunov exponents at the soft edge follow GUE statistics, associated with quantum chaotic behaviour.
 {We filled the gap in analytically describing the local spectral statistics 
in the bulk and at the edge of the spectrum
in the critical regime $M\propto N$, 
in deriving two limiting kernels 
that
interpolate between the 
deterministic and GUE regimes.
Interestingly, the transition between these two phases still exists for $M\ll N$ albeit not the whole spectrum lies in one or the other phase. The critical regime shrinks to a narrow scale about the hard edge.
}

{Our findings agree with Dyson's BM with picket fence 
initial conditions~\cite{KurtBrown}, showing the universality of our results.
Here, we want to underline that this agreement is far from obvious since products
of matrices cannot be easily expressed into sums of other independent objects like the sums of their logarithms.
Indeed, this essential difference to scalars can be noticed by the validity of the map to Dyson's BM, which only holds locally and not for the whole spectrum. For instance at the hard edge,
the deterministic behaviour for the smallest Lyapunov exponents persists, regardless of what relation $N$ and $M$ have.}

{
The relation between transfer matrices and the Dyson BM studied in~\cite{KurtBrown}
has been already pointed out in~\cite{IpsenSchomerus} for a specific choice of products of random matrices
modelling the stochastic process~\eqref{stochproc}. Therein, the product of matrices perturbing the unit matrix has
been analysed which yields the DMPK equation in the limit of large $M$.
Therefore, we believe that our interpolating regime is related to
the Anderson transition, 
also resulting from the DMPK equation~\cite{D,MPK,KlausFrahm,Beenakker,IpsenSchomerus}.}

Since our model is analytically solvable  
for any finite $N$ and $M$,
one can also 
address
the nearest neighbour spacing distribution and the distribution of the largest Lyapunov exponent. The latter has important consequences for the stability of the system  and its Kaplan-Yorke dimension~\cite{ky}.
Another direction 
of investigation 
would be the analysis of the microscopic statistics of the 
complex eigenvalues 
of the product matrix. For the spectral radius first investigations were done~\cite{JiangQi,GuiQi,LiQi}, 
finding a transition, too.

\begin{acknowledgments}
We acknowledge support by the German research council (DFG) 
via
CRC 1283: ``Taming uncertainty and profiting from randomness and low regularity in analysis, stochastics and their applications" (GA, MK)  and by the AGH UST statutory tasks
No. 11.11.220.01/2 within subsidy of the Ministry of Science and Higher Education (ZB). 

We thank D.-Z. Liu and D. Wang for sharing their results about the same double scaling limit at the soft edge derived independently~\cite{LWW}, and are grateful for D.-Z. Liu's comments on a preprint version of this letter. Moreover, we are indebted to M. Duits for pointing out a possible relation to~\cite{KurtBrown}, and to 
J.J.M. Verbaarschot for 
confirming this 
for the microscopic density in the bulk, by applying the Poisson summation formula.  Finally, we also appreciate fruitful discussions with J. Ipsen on the DMPK equation.
\end{acknowledgments}

\end{document}